\documentstyle[epsfig,12pt]{elsart} 
\begin{document}
\newcommand{\p}{\partial}
\newcommand{\ls}{\left(}
\newcommand{\rs}{\right)}
\newcommand{\beq}{\begin{equation}}
\newcommand{\eeq}{\end{equation}}
\newcommand{\beqa}{\begin{eqnarray}}
\newcommand{\eeqa}{\end{eqnarray}}
\newcommand{\bdm}{\begin{displaymath}}
\newcommand{\edm}{\end{displaymath}}

\begin{frontmatter}
\title{On the Lorentz structure of the symmetry energy}
\author[catania]{T. Gaitanos},
\author[catania]{M. Di Toro\thanksref{dit}},
\author[gsi]{S. Type l},
\author[catania]{V. Baran\thanksref{bar}},
\author[tuebingen]{C. Fuchs},
\author[texas]{V. Greco\thanksref{gre}}
and
\author[muenchen]{H.H. Wolter}

\address[catania]{Laboratori Nazionali del Sud INFN, I-95123 Catania, Italy\\
 Physics \& Astronomy Dept., Univ. of Catania }
\address[gsi]{Gesellschaft f\"ur Schwerionenforschung, D-64291 Darmstadt, 
Germany}
\address[tuebingen] {Institute f\"ur Theoretische Physik, Universit\"at
 T\"ubingen, D-72076 T\"ubingen, Germany}
\address[texas]{Cyclotron Institute, Texas A\& M University, College Station,
 USA}
\address[muenchen]{Sektion Physik, Universit\"at M\"unchen, 
D-85748 Garching, Germany}  

\thanks[dit]{ditoro@lns.infn.it}
\thanks[bar]{On leave from NIPNE-HH and Bucharest University, Romania}
\thanks[gre]{Supported by a INFN Post-Doc Fellowship}
\begin{abstract}
We investigate in detail the density dependence of the symmetry energy 
in a relativistic description by decomposing the 
iso-vector mean field into contributions with different Lorentz covariant 
properties. We find important effects of the iso-vector, scalar  
channel (i.e. $\delta$-meson like) on the high density behavior 
of the symmetry energy.
Applications to static properties of 
finite nuclei and to dynamic situations of heavy ion collisions 
are explored and 
related to each other. The nuclear structure studies show only moderate 
effects 
originating from the virtual $\delta$ meson. At variance, in heavy 
ion collisions one 
finds important contributions on the reaction dynamics arising from the 
different 
Lorentz structure of the high density symmetry energy 
when 
a scalar iso-vector $\delta$ field is introduced. Particularly interesting
is the related neutron/proton effective mass splitting for nucleon transport 
effects and for resonance and particle production around the threshold.
We show that the $\delta$-like channel 
turns out 
to be essential for the production of pions,  
when comparing with 
experimental data, in particular for high momentum selections.
\end{abstract}
\end{frontmatter}
\section{Introduction}

Heavy ion collisions at relativistic energies from 
$0.1$ up to $1-2~AGeV$ offer the possibility to access the equation 
of state ($EOS$) of nuclear matter under extreme conditions of density,
 isospin  
and temperature \cite{ritter97}. This knowledge is important in 
understanding many interesting
astrophysical phenomena such as the physical mechanism of supernovae 
explosions and the neutron star structure and cooling. 

During the last 
three 
decades one has attempted in many ways to investigate the properties of 
highly excited hadronic matter. In phenomenological models of a 
relativistic quantum field theory, like the Relativistic Mean Field
($RMF$) model of 
Quantum-Hadro-Dynamics ($QHD$)\cite{qhd},
 a significant decrease of the effective nucleon mass 
with density was found to be essential for a reliable description of 
infinite nuclear matter and finite nuclei. More microscopic 
studies, the so-called Dirac-Brueckner-Hartree-Fock ($DBHF$) approach,
based on realistic nucleon-nucleon ($NN$) interactions in the 
One-Boson-Exchange ($OBE$) model, see 
refs.\cite{mach89,brmach90,mal92,mDBHF},
 also predict a strong decrease 
of the effective nucleon mass. Similar effects have been found 
for particles with strangeness such as kaons \cite{ko99,giessen,fuc01}. 
The results can be represented by saying 
that the effective nucleon-meson vertices or 
coupling functions 
exhibit a density dependence and, for energies higher than 
the Fermi energy, an additional momentum dependence \cite{mDBHF}. As a 
consequence the $EOS$ predictions of different models 
show quite variable density 
dependences for densities 
higher than saturation, characterized by referring to a hard or soft $EOS$, 
which have been tested in heavy ion collisions \cite{EOS}. 

So far asymmetric nuclear matter has been only poorly investigated,
but certainly the same considerations appear for the density dependence
of the symmetry term which are directly related to the Lorentz structure
of the effective fields in the isovector channel and to their weights
(the coupling functions) in different baryon density regions.
Around the saturation point the $a_4$ parameter of the Weizs\"{a}ecker mass 
formula contains the leading information (in a sense equivalent to the $a_1$
parameter for the isoscalar part). Actually, there is quite a wide range of
extracted values, roughly from $28~to~36MeV$, depending on the way the
surface effects are accounted for. This clearly shows that even to
reproduce the equilibrium $EOS$ point we need a better knowledge of
the density dependence of the symmetry term since the slope (the $symmetry~
pressure$) and the curvature (the $symmetry~compressibility$) around
$\rho_0$ are also playing an important role in reproducing nuclear masses.
 A related sensitive observable appears to be the difference
between neutron and proton r.m.s. radii in charge asymmetric nuclei.
The wide range of predictions made by different relativistic 
(and non-relativistic
Skyrme-$HF$) effective interactions has been nicely presented
 recently
in the refs. \cite{tybr01,fur02}. Other independent information on the
symmetry energy in dilute matter, below saturation, have been suggested
from heavy ion collisions in the Fermi energy domain, in particular
from the $isospin~distillation$ mechanism in fragment production, 
 \cite{bar02} and refs. therein, and from the isospin content of fast 
particle emissions, \cite{bao00}.   

In standard $RMF$ models the iso-vector part of the mean field is 
described in terms of a constant vector, isovector, $\rho$-coupling giving 
an almost  linear density dependence of the symmetry energy. 
Recently the importance of
a contribution of the scalar isovector channel, the virtual $\delta[a_0(980)]$
meson, within the same $RMF$ framework, has been stressed \cite{liu02,gre03}.
This mostly affects the high density behavior, as we will show in the 
following. 
We note that the $\delta-$channel usually has not been considered 
before just on the basis of the weak contribution to the free 
Nucleon-Nucleon interaction. 
But in the spirit of the
$Effective~ Field~ Theory$  as a relativistic $Density~ Functional~ Theory$,
(the $EFT/DFT$ framework, see \cite{fur00,fur03}), the relevance of 
this channel
could be completely different in nuclear matter, due to medium and
many-body effects. 
We like to note that very recently , see the conclusions of ref.\cite{fur02}, 
the $\delta$-channel has been reconsidered
as an interesting improvement of covariant approaches, in the framework
of the $EFT/DFT$ philosophy. One of the main tasks of
our work is just to try to select the dynamical observables more sensitive
to it. 

In the Density Dependent Hadronic ($DDH$) 
field theory \cite{lenske}, the detailed density dependence
of the nucleon-meson couplings has been taken either directly
from $DBHF$ calculations \cite{hole01,sch01,mali02} or has been fitted
to data of nuclear matter and finite nuclei \cite{typel}. 
Certainly the correct microscopic approach should be to derive the coupling 
functions, in a $QHD$ mean field scheme, from Dirac-Brueckner-Hartree-Fock
calculations. Several attempts have been performed so far, see refs.
\cite{hole01,sch01,mali02}, but the results do not agree and appear to
be model dependent, in particular for the isovector channel. 
We will discuss such point in more detail below.

Aim of the present work is to try to shed some light on this intricate 
puzzle of the baryon density dependence of the symmetry term, in particular
in the high density region. For densities beyond saturation one needs 
extrapolations which can only be investigated in dynamical situations of 
heavy ion collisions at intermediate energies, where hot and dense 
asymmetric nuclear matter is formed. 
Indeed, the study of the high density symmetry energy 
is an object of recently started investigations \cite{bali,grefl,lili,kofra}. 
Their motivation has been the 
planning of new experimental heavy ion facilities with neutron-rich 
radioactive beams with data available in the near future. 

Here we will actually explore the properties of asymmetric nuclear matter
in different nuclear systems, i.e. finite nuclei and heavy ion collisions. 
For our studies we apply various versions of $QHD$
 models, 
 in the Relativistic Mean Field ($RMF$) approach, which 
differ 
in the treatment of the effective meson-nucleon vertices. In the 
standard  $RMF$ models the nucleon-meson 
couplings are given by constants fitted to nuclear matter saturation 
properties. Some non-linear contributions for the 
meson fields are considered, $NL-RMF$ approaches, which imply definite 
density dependences in the
model. The $RMF-DDH$ theory uses the so-called 
Vector Density Dependent ($VDD$) parametrizations \cite{lenske,typel}
in terms of directly density dependent effective coupling functions.
In this case there is no need of non-linear terms.

In this context relativity is important because 
it naturally explains essential nuclear matter properties such as the 
cancellation of large Lorentz vector and scalar fields, in the isoscalar
channel, leading to a small 
value of $16~MeV$ for the binding energy per nucleon and, simultaneously, to 
a strong spin-orbit potential. Within a covariant formulation one is also 
able to better understand  
the origin of the repulsive and attractive contributions to the effective $NN$ 
interaction. The separation of the 
total nuclear 
mean field potential in terms of its Lorentz components is 
of particular importance when studying dynamical situations of 
heavy ion collisions at relativistic energies \cite{grefl,rela,fuc96}. 

In the $RMF$ picture 
the nuclear mean field is given by the exchange of mesons with different 
Lorentz properties. The iso-scalar part which describes the $EOS$ of symmetric 
nuclear matter is mainly characterized by an attractive scalar 
field ($\sigma$-meson) 
and repulsive vector field ($\omega^{\mu}$-meson) both with 
contributions of the same order $400-500~MeV$ to the total energy per nucleon. 
The corresponding masses, fixed as $m_{\sigma} \sim 550~MeV$ and 
$m_{\omega}=783~MeV$, are important in finite nuclei
calculations. 
One should note the importance of both these mesons in describing 
the saturation 
properties of symmetric nuclear matter. 

A covariant treatment of the problem is also helpful for the 
understanding 
of the properties of asymmetric nuclear and neutron matter. As in the 
iso-scalar 
case, the iso-vector part of the nuclear mean field, which is responsible 
for the 
density dependence of the symmetry energy, can again be 
characterized by iso-vector mesons 
with different Lorentz properties. These are the 
iso-vector, vector field ($\vec{\rho}^{\mu}$-meson) and the iso-vector, scalar 
field ($\vec{\delta}$-meson), with masses $m_{\rho}=769~MeV$ 
 ($763~MeV$ in $VDD$ models) and 
$m_{\delta}=980~MeV$. Due to less available experimental information 
several possibilities have been proposed to describe the density 
dependence of 
the symmetry energy. 

In contrast to the iso-scalar case, since here we do not have the stringent
saturation condition of balancing attractive (scalar) and repulsive (vector)
contributions, one can apply 
either only the $\rho$-meson, 
or both, the $\rho$- and $\delta$-mesons. Because of the different 
Lorentz properties 
of the mesons one expects different density behavior of the symmetry 
energy especially 
at high densities. The argument is very simple and clearly discussed in the 
refs. \cite{liu02,gre03}. Vector and scalar effective fields will contribute
to the symmetry term with quantities proportional to baryon and scalar 
densities, respectively. As a consequence the main differences should appear
at baryon densities above saturation.
Of course such direct interplay between $\rho$ and $\delta$ contributions
to the symmetry energy can be reproduced in the symmetry term just by
introducing density dependent isovector/vector couplings without the need
of a new virtual (scalar) meson in the isospin channel. 
The scalar nature of the
$\delta$-meson, however, also leads to a new effect, the splitting of the
neutron and proton effective masses, with interesting consequences on
nucleon transport properties and on particle and resonance production
in the reaction dynamics at high baryon and isospin densities.
In this work we will focus our attention on these new features of a 
isovector/scalar coupling in realistic collision simulations and
on the related most sensitive experimental observables. Charged pion yields 
appear to be a suitable quantity to measure in order to see such effects. 

As we shall show, finite nuclei studies 
cannot separate easily
between the effects originating from the different iso-vector 
mesons for densities mainly
below saturation, see also \cite{typel}. Therefore, relativistic heavy 
ion collisions 
represent a unique tool 
to study the properties of highly compressed asymmetric nuclear matter using 
radioactive beams. 

The paper is organized as follows. In Section 2 we discuss 
asymmetric 
nuclear matter within the $RMF$ theory. 
The models are then applied to static cases of finite nuclei (Sec.3) and to 
dynamical ones 
of heavy ion collisions (Secs.4, 5). The application to finite nuclei 
is important to demonstrate 
the reliability of the models in this sector, and the role of the 
$\delta$-meson 
will be particularly emphazised. Then we describe briefly the theoretical 
treatment 
and numerical procedure of heavy ion collisions. The transport 
theoretical analysis 
is mainly applied here to particle production. 
In particular we focus the attention on charged pion yields,
 with rapidity and transverse momentum selections (Sec.5).  
A comparison with experiments is performed, when data are available, also in a 
preliminary status. Final remarks and an outlook follow in the last Section.

\section{Equation of state of asymmetric nuclear matter}
The starting point of a covariant quantum field theory is the 
Lagrangian of an interacting many body system of baryons and 
mesons. The latter determines the interaction between baryons 
in terms of nucleon-meson coupling vertices. Within a Non-Linear  
Relativistic Mean Field ($NL-RMF$) approximation the baryons 
are given by Dirac 
spinors and the mesons are described by classical field equations 
\begin{eqnarray}
& & [\gamma_{\mu}i\partial^{\mu} - g_{\omega}\omega_{0}\gamma^{0} - 
g_{\rho}\gamma^{0}\tau_{3}\rho_{0} - 
(M - g_{\sigma}\sigma - g_{\delta}\tau_{3}\delta_{3})]\Psi = 0
\label{dirac}\\ 
& & m_{\sigma}^2\sigma + a\sigma^2 + b\sigma^3 = 
g_{\sigma}< \hat{\overline\Psi}\hat{\Psi} > 
= g_{\sigma}\rho_s
\label{gordon}\\ 
& & m_{\omega}^2\omega^{\mu} = 
g_{\omega}< \hat{\overline\Psi}\gamma^{\mu}\hat{\Psi} > 
= g_{\omega} j^{\mu}
\label{proca}\\
& & m_{\rho}^2\rho^{0} = 
g_{\rho}< \hat{\overline\Psi}\gamma^{0}\tau_{3}\hat{\Psi} > 
= g_{\rho} \rho_{3}
\label{rhomeson}\\
& & m_{\delta}^2\delta_{3} = 
g_{\delta}< \hat{\overline\Psi}\tau_{3}\hat{\Psi} > 
= g_{\delta} \rho_{s3}
\label{deltameson}
\quad .
\end{eqnarray}
The nuclear system is characterized by 
a baryonic quantum field and a surrounding 
background of classical mesonic fields, the iso-scalar, scalar $\sigma$-, the 
iso-scalar, vector $\omega$-, the iso-vector, scalar $\delta$- and 
the iso-vector, vector $\rho$-mesons. The isospin vector and 
scalar densities are given 
by $\rho_{3}=\rho_{p}-\rho_{n},~\rho_{s3}=\rho_{sp}-\rho_{sn}$, 
respectively, $\rho_{p,n}$ being the proton and neutron densities. 
The iso-scalar $\sigma$ field contains non-linear contributions.
In $DDH-RMF$ approaches the above equations will be modified
for the presence of rearrangement terms in the nucleon self-energies
and for the lack on non-linear terms, see refs. \cite{lenske,typel}.
 
The energy-momentum tensor within the same non-linear $RMF$ approach is 
given by 
($f_{i} \equiv (g_{i}/m_{i})^{2}$) :
\begin{eqnarray}
T^{\mu\nu}(x) & = & t^{\mu\nu} + f_{\omega} j^{\mu}j^{\nu} + 
f_{\rho}j^{\mu}_{3}j^{\nu}_{3}                    
\nonumber\\
& + & \frac{1}{2}g^{\mu\nu} 
                    \left[ 
                        m_{\sigma}\sigma^{2} + 2U(\sigma) + 
                        f_{\delta}\rho_{s3}^{2} - 
                        f_{\omega}j^{\alpha}j_{\alpha} - 
 f_{\rho}j^{\alpha}_{3}j_{3\alpha} 
                    \right]
\label{energy}
\qquad ,
\end{eqnarray}
with $t^{\mu\nu}$ the kinetic part, 
$t^{\mu\nu}(x) \equiv 
 i< \hat{\overline\Psi}\gamma^{\mu}\partial^{\nu}\hat{\Psi} >$, 
   and $U(\sigma)$ the non-linear 
contributions of the iso-scalar $\sigma$-meson 
$U \equiv \frac{1}{3}a\sigma^{3}+\frac{1}{4}b\sigma^{4}$. 
The effective mass is now different for protons and neutrons due to the 
appearance of the iso-vector, scalar  $\delta$-meson \cite{npmass}:
\begin{equation}
m^{*}_{i}=M-g_{\sigma}\sigma \pm g_{\delta}\delta_{3}
\quad\mbox{(- proton, + neutron)}\quad
\label{effmass}
\quad .
\end{equation}
\nopagebreak
\begin{table}[t]
\begin{center}
\begin{tabular}{|l|c|c|c|c|c|c|c|c|c|}
\hline\hline 
       & $f_{\sigma}$ ($fm^2$)   & $f_{\omega}$ ($fm^2$) & $f_{\rho}$ ($fm^2$)
     & $f_{\delta}$ ($fm^2$)     & A ($fm^{-1}$) & B &  \\ 
\hline\hline
   $NL\rho$          &     15.60        &     10.50     & 1.10 &    0.0     &
  0.015     &    -0.004  \\ 
\hline
   $NL\rho\delta$ &     15.60        &     10.50    & 3.15 &    2.4     &
  0.015     &    -0.004  \\ 
\hline
   $NL3$                 &     15.73        &     10.53    & 1.34 &    0.0     &
   0.01     &    -0.003   \\ 
\hline\hline
\end{tabular}
\end{center}
\vskip 0.5cm
\caption{\label{table1} 
Coupling parameters in terms of $f_{i} \equiv 
(\frac{g_{i}}{m_{i}})^{2}$ 
for $i=\sigma,~\omega,~\rho,~\delta$, $A \equiv \frac{a}{g_{\sigma}^{3}}$ and 
$B \equiv \frac{b}{g_{\sigma}^{4}}$ for the non-linear $RMF$ models 
using the $\rho$ and 
both, the $\rho$ and $\delta$ mesons for the characterization of the 
isovector mean 
field in comparison with the $NL3$ model. See text. }
\vskip 0.5cm
\end{table}


For the investigation of asymmetric nuclear matter and neutron matter we 
introduce the asymmetry parameter 
$\alpha=\frac{\rho_{n}-\rho_{p}}{\rho_{n}+\rho_{p}}$ which describes the 
relative ratio of the neutron to proton fraction of the nuclear matter. 
The asymmetric nuclear matter calculations can be done by solving 
selfconsistently the  
Eqs. (\ref{dirac}-\ref{deltameson},\ref{effmass}). The parameters 
of the model, fixed to nuclear matter saturation properties, are 
given in the Table \ref{table1} and compared to the widely used $NL3$ 
parametrization \cite{ring,nl3}. In order to isolate the effects in 
the isovector channel
we have chosen on purpose very similar parameters in the isoscalar part.
At saturation in symmetric matter we have a density $\rho_{sat}=0.148fm^{-3}$,
a nucleon effective mass $m^*_N=0.6M$ and a compressibility 
$K_{NM}=271.7MeV$. In $NL3$ the symmetry parameter is $a_4=37.4MeV$
while in our $NL\rho$, $NL\rho\delta$ choices we fit a value
$a_4=30.5MeV$.

In all the above $NL-RMF$ schemes the nucleon-meson couplings
are given by constant 
values, see Table \ref{table1}. 
We also want to compare to $RMF$ models based on the Density Dependent 
Hadronic ($DDH$) field approach \cite{lenske}, since they have been
shown to reproduce well nuclear matter and finite nuclei  
\cite{hole01,typel,ringdd}. 
 We will follow here in particular the 
parametrization of ref.\cite{typel}, where the density dependence has been 
fitted to finite nuclei, and which makes a prediction for the isovector
channel at high density. The symmetry energy at saturation is chosen 
as $a_4\sim33.4MeV$, see also ref.\cite{ringdd}.
The corresponding properties for symmetric matter are: density
$\rho_{sat}=0.153fm^{-3}$,
nucleon effective mass $m^*_N=0.55M$ and compressibility 
$K_{NM}=240.0MeV$. So the interactions with density dependent couplings
used in this work are softer with respect to the $NL-RMF$ models. We will see
that this will affect the particle production yields. 
 
However, the work of ref.\cite{typel} made no predictions for the
$\delta$-like field, since there was not much sensitivity to it in finite
nuclei, as expected (see Sect.1) and as we will also see here later in Sect.3.
To test this part in heavy ion collisions we use three variants of these
$DDH$ models. $DDH\rho$ is just the $VDD$ parametrization of ref.\cite{typel},
with one a $\rho$-like isovector field with an exponential density dependence
of the coupling. In the models $DDH3$ we use the functional form of the
vertex density dependences for the $\rho$- and $\delta$-fields as extracted
from the $DBHF$ calculations of ref.\cite{lenske2}.
In $DDH3\rho$ we use only the $\rho$-field and normalize to the
$a_4\sim33.4MeV$ value at saturation, as in the $DDH\rho$ ($VDD$) case.
In the $DDH\rho\delta$ we use $\rho$- and $\delta$-fields and normalize,
see following, as to obtain the same symmetry energy at saturation.

\begin{figure}[t]
\unitlength1cm
\begin{picture}(10.,9.0)
\put(0.0,0.0){\makebox{\epsfig{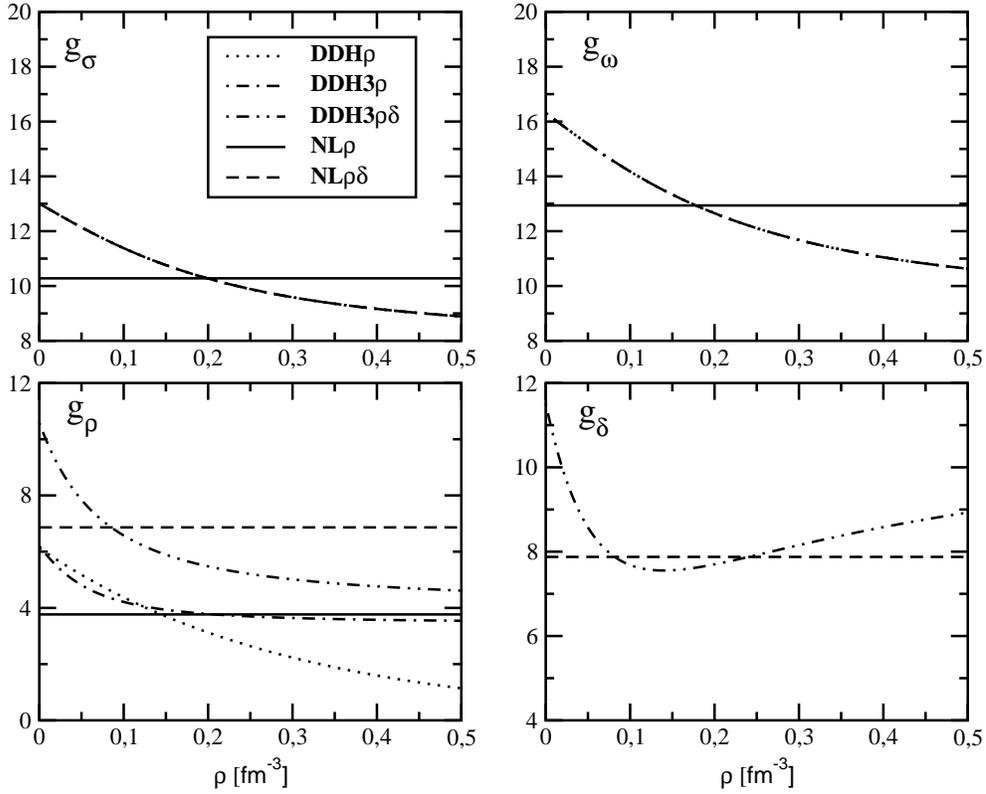}}}
\end{picture}
\caption{
Density dependence of the different coupling functions as indicated 
for the relativistic mean field models used here 
(see text for details).
}
\label{fig1}
\end{figure}

The density dependence of the couplings, for all channels, for all the
tested models is shown in Fig.\ref{fig1}. In the $NL-RMF$ models the
couplings are constant, for the $DDH-RMF$ model they are chosen as
discussed above.
Note, in particular, the unusual density dependence of the
$\delta$-like vertex and the attempt to reproduce some average constant
value in the corresponding $NL\rho\delta$ parametrization.
Finally we like to remind that, due to the non-linear
self interaction of the $\sigma$-meson, in the $NL-RMF$ models the
effective $g_{\sigma}$ will also show a density dependence.

The symmetry energy $E_{sym}$ is defined from the expansion of the 
energy per nucleon $E(\rho_B,\alpha)$ in terms of the asymmetry parameter
(we use $\rho_B$ for the baryon 
density in order to avoid confusion with the $\rho$ meson):

\begin{equation}
E(\rho_B,\alpha) = E(\rho_B) + E_{sym}(\rho_B)\alpha^{2}
+{\cal O}(\alpha^{4})+ \cdots 
\label{esym1}
\end{equation}
with the abbreviation 
\begin{equation}
E_{sym} = \frac{1}{2} 
\frac{\partial^{2}E(\rho_B,\alpha)}{\partial \alpha^{2}}|_{\alpha=0}
=
\frac{1}{2} \rho_B
\frac{\partial^{2}\epsilon}{\partial \rho_{3}^{2}}|_{\rho_{3}=0}
\label{esym2}
\quad .
\end{equation}
From the definition of the energy momentum tensor (\ref{energy}) one obtains 
for the symmetry energy \cite{liu02,gre03}:
\begin{equation}
E_{sym} = \frac{1}{6} \frac{k_{F}^{2}}{E_{F}} + 
\frac{1}{2}
\left[ f_{\rho} - f_{\delta}\left( \frac{m^{*}}{E_{F}} \right)^{2}
\right] \rho_{B}
\label{esym3}
\quad .
\end{equation}
where $E_{F}=\sqrt{k_F^2+{m^*}^2}$.

It is seen that including the scalar iso-vector $\delta$-meson the 
bulk asymmetry parameter is given by the combination 
$[ f_{\rho} - f_{\delta}( \frac{m^{*}}{E_{F}})^{2}]$ of the 
repulsive vector ($\rho$) and attractive ($\delta$) iso-vector couplings. 
We thus have to increase the $\rho$-meson coupling when including the 
$\delta$-meson in the iso-vector part of the equation of state in order 
to reproduce the same bulk asymmetry parameter $a_{4}$. This is a very
general feature, not depending on the 
model, that we will see to be important in the dynamical simulations.
Another $\delta$-effect, of interest for transport properties, will be
the splitting of the neutron/proton effective masses \cite{liu02,gre03},
 see Eq.(\ref{effmass}).

From Eq.(\ref{esym3}) we note the moderate role of the 
scalar, iso-vector $\delta$-meson at low densities, and therefore 
on properties of finite nuclei (see below), 
as already pointed out in \cite{typel}.  It is 
important to realize the different Lorentz structure of the iso-vector 
mesons. The vector $\rho$-meson contribution depends linearly on the 
baryon density whereas 
the scalar $\delta$-meson part is proportional to the scalar density 
or correspondly to 
the factor $(m^{*}/E_{F})^{}$, decreasing at high baryon density.

\begin{figure}[t]
\unitlength1cm
\begin{picture}(10.,9.0)
\put(0.0,0.0){\makebox{\epsfig{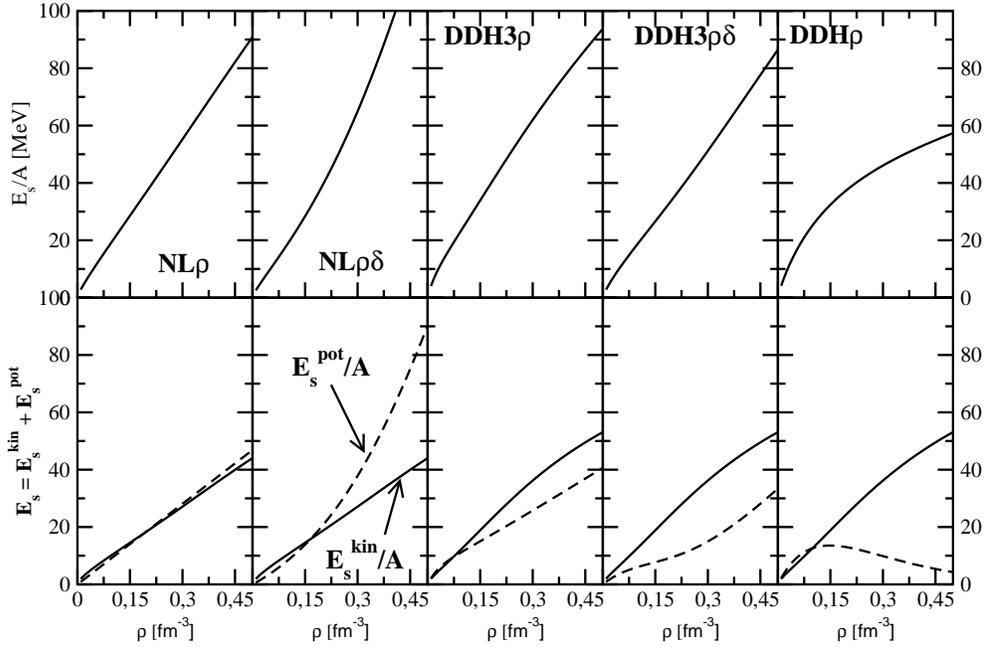}}}
\end{picture}
\caption{Density dependence of the symmetry energy, Eq. (\protect\ref{esym3}), 
for the same models as in Fig.~\protect\ref{fig1}. The lower part shows 
separately the kinetic (solid) and potential (dashed)
contributions to the total symmetry 
energy.
}
\label{fig2}
\end{figure}

In Fig. \ref{fig2} the density dependence of the symmetry energy, 
Eq. (\ref{esym3}), for the different models used here is shown. 
The kinetic and the potential contributions to the 
total symmetry energy can be also seen separately on the bottom. 
Within the Walecka models ($NL\rho$ and $NL\rho\delta$)) one 
observes a similar behavior of $E_{sym}$ at densities below saturation. 
With increasing baryon density $\rho_{B}$, however, the differences arising 
from the presence of the $\delta$-meson in the iso-vector channel become 
more pronounced, as expected. This is due to the saturation of the scalar 
density (the factor $( \frac{m^{*}}{E_{F}} )^{2}$ of Eq.(\ref{esym3})) 
for high densities and, 
on the other hand, to the enhanced value of the vector $\rho$-meson 
contribution
(needed to reproduce the same $a_4$, see Table 1) 
which then grows linearly with baryon density. 

In the $DDH$ models the high density influence 
of the $\delta$-meson on the iso-vector part of the equation of state 
is more involved. We have now a rapid decrease 
with density of the $\rho-$coupling function while the
increase of the $\delta-$coupling above saturation almost counteracts 
the relativistic
scalar density effect, see Fig.\ref{fig1}.
As we can see from the two $DDH3$ columns of Fig.\ref{fig2}, 
with the inclusion of the $\delta$-meson the symmetry energy is not 
much changed for baryon densities 
higher than saturation and actually can become even softer.
The Lorentz structure of the isovector contributions is now different,
and moreover we have an important splitting of the $n/p$ effective
masses. We will show that 
from particle
production it may be possible to trace back these fine relativistic effects,
although of course difficult.
 The 
density behavior of symmetry energy for the the $DDH\rho$ model, 
(last column of Fig.\ref{fig2}), can be regarded as 
very soft originating from a relatively strong density decrease 
of the $\rho$-meson coupling, see Fig.\ref{fig1}. This is particularly evident
in the potential contribution (dashed line on the bottom-right).
From Fig.\ref{fig2} the importance of relativistic features in 
the understanding 
of the high density behavior of the symmetry energy is evident. 

In contrast to the iso-scalar part of the mean field, which 
has been extensively investigated in the last three decades theoretically and 
experimentaly \cite{ritter97,dani}, 
there is not enough empirical information about the high density behavior 
of the symmetry energy. Heavy ion collision experiments with 
radioactive beam facilities 
have recently been started or are being planned. In the following 
section we 
investigate first in finite nuclei the Lorentz structure of the 
iso-vector channel in the framework of the models discussed here. 
In the next section we will then pass to heavy ion collisions at 
intermediate energies,
 where higher densities can be reached and relativistic effects can
be enhanced. We 
will discuss constant coupling Walecka-type models 
($NL\rho$ and $NL\rho\delta$)) as a kind of reference 
to clearly see the physics of the relativistic effects of different
isovector contributions. We will compare then with the more sophisticated 
$DDH\rho$ and $DDH3\rho$, $DDH3\rho\delta$ models in order make
predictions for the main observables in experiments.

\section{Finite nuclei (static case)}

In this Section we focus on the role of the iso-vector, scalar 
$\delta$-meson in the properties of finite nuclei and, particularly, 
we compare with other models widely used in nuclear structure 
studies. 
For the description of finite nuclei we solve the equations of motion 
for nucleons and mesons, Eqs. (\ref{dirac}-\ref{deltameson}), including 
the gradient terms in the classical meson field equations in a 
selfconsistent scheme. 
In case of density dependent coupling functions also additional 
rearrangement contributions of the iso-scalar and iso-vector classical 
fields (not shown in Eqs.(\ref{dirac}-\ref{energy})) are explicitely 
taken into account \cite{typel,lenske}.

\begin{figure}[t]
\unitlength1cm
\begin{picture}(10.,9.0)
\put(0.0,0.0){\makebox{\epsfig{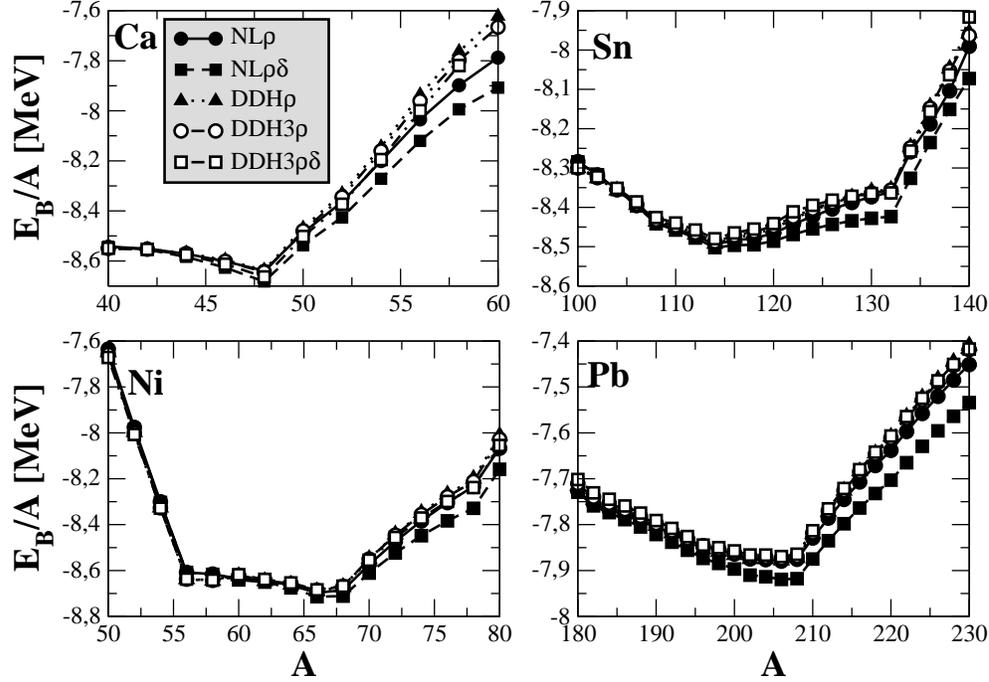}}}
\end{picture}
\caption{
Mass dependence of the binding energy per nucleon for $Ca,~Ni,~Sn$ and 
$Pb$ isotopes for the different models.
}
\label{fig3}
\end{figure}
\begin{figure}[t]
\unitlength1cm
\begin{picture}(10.,9.0)
\put(0.0,0.0){\makebox{\epsfig{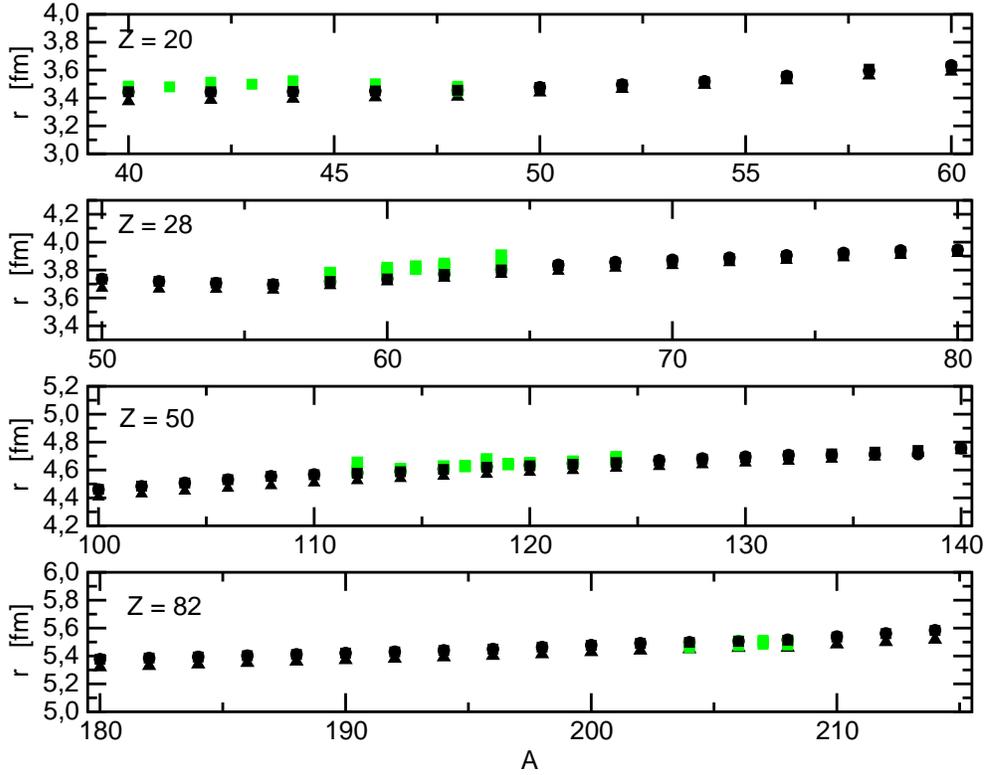}}}
\end{picture}
\caption{Mass dependence of charge proton radii for $Ca-$, $Ni-$, $Sn-$ 
and $Pb-$isotopes: (grey squares) exp. data, (full triangles) $DDH\rho$, 
(open triangles) $NL3$, (full circles) $NL\rho$) and 
(full squares) $NL\rho\delta$). 
}
\label{fig4}
\end{figure}
Fig. \ref{fig3} displays the mass dependence of the 
binding energy per nucleon for different isotopes, starting 
from the most symmetric case ($Ca$) to the most asymmetric one ($Pb$).
The results for all models are close together but 
the differences originating from the $\delta$-meson become 
more pronounced with increasing neutron to proton 
asymmetry. This is particularly evident for the constant coupling $RMF$ 
models, when comparing $NL\rho$ and $NL\rho\delta$ results.


We recall from Fig.\ref{fig2} that in the region around and below 
saturation, tested in finite nuclei,  
no significant differences between the models are seen in the 
symmetry energy. Only between the two extreme cases of $NL\rho\delta$ 
and $DDH\rho$ there is a visible difference in the low density regime which 
also leads to effects in the binding energy per nucleon 
for highly asymmetric finite systems. Generally, the binding energy becomes 
larger with the $\delta$-meson included in the iso-vector mean field 
because here the attractive mean field is stronger (smaller) for neutrons 
(protons), with an observed net effect for neutron-rich nuclei. 
The comparison with the widely used models $DDH\rho$ \cite{typel} and 
$NL3$ \cite{ring,nl3} (here essentially given by the $NL\rho$ points), 
which nicely
describe finite nuclei properties, indicates the consistency 
of our parametrizations, that we will then extend to the 
description of heavy ion collisions. 

\begin{figure}[t]
\unitlength1cm
\begin{picture}(10.,9.0)
\put(0.0,0.0){\makebox{\epsfig{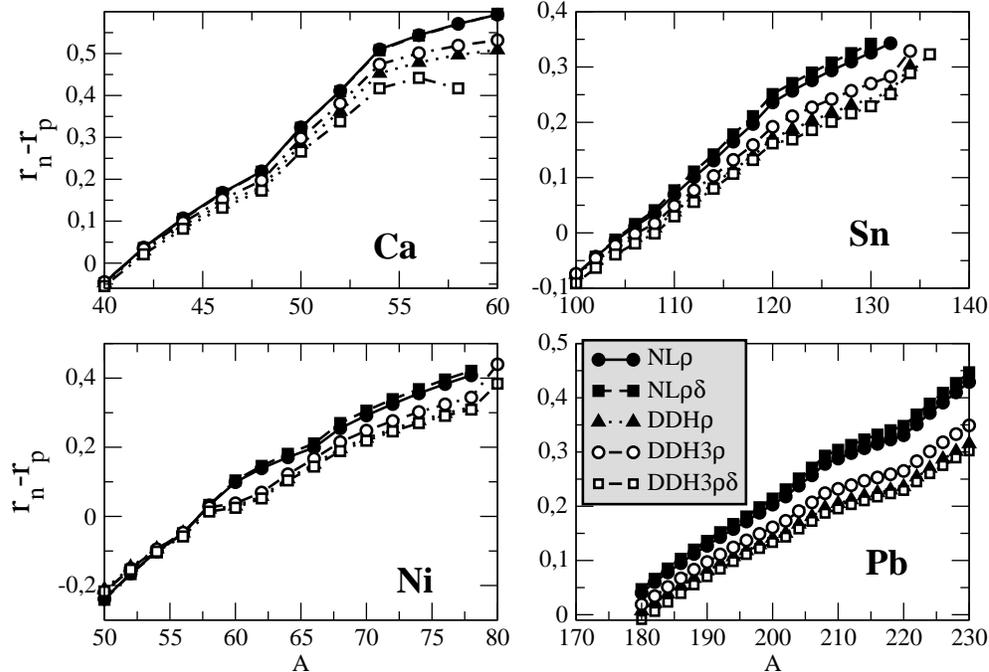}}}
\end{picture}
\caption{
Mass dependence of the difference of the neutron to proton radii 
for the same isotopes and models as in Fig.~\protect\ref{fig3}.
}
\label{fig5}
\end{figure}

Fig.~\ref{fig4} shows the charge proton radii for different isotopes, 
 as indicated. First, one observes moderate differences between the 
different models, since the proton density profiles are very 
similar, as expected. All the models can reproduce the experimental 
values, as an important issue.
 
One should expect larger effects of the internal 
structure of the symmetry 
energy by looking at the difference between neutron and proton radii,
which is more sensitive to slope and curvature of the symmetry energy around
saturation \cite{tybr01,fur02}. 
This is shown in Fig.~\ref{fig5} for the same models and systematics. 
As in the previous figures, the influence of the $\delta$-meson is again 
not much pronounced in the $NL-$ and $DDH-$models, because of the 
similar density 
behavior of the symmetry energy at low densities. We can see a systematic
smaller $r_n - r_p$ for neutron-rich nuclei, i.e. a reduced neutron 
diffusivity, in the case of
$DDH$ parametrizations, due to the larger $\rho$-(symmetry energy)
 repulsion seen by neutrons
on the surface (at lower densities), see Fig.\ref{fig1}. This seems to be 
in better agreement with data, see \cite{fur02}. We note that this
result is present also when the $\delta$ channel is added, see the
open squares of the $DDH3\rho\delta$ interaction, in particular for the
largely discussed $^{208}Pb$ case. 
 
At variance the  slightly smaller symmetry repulsion for neutrons 
at densities below 
$\rho_0$ in the $NL\rho\delta$ parametrization can consistently 
account for the larger neutron diffusion and the overall larger binding
energies for very neutron-rich nuclei, Fig.\ref{fig3}.

This discussion confirms the relatively small effect of a more extended
relativistic structure of the symmetry term in the low density range,
tested in finite nuclei structure calculations, see also \cite{typel}.  
Similar results have been observed in the spin-orbit splitting of 
single particles 
levels, which are not shown here. 

Generally the properties of finite nuclei can be well reproduced in the 
$NL$ models and are close to the results of more realistic $DDH$ approaches 
\cite{typel}. This is an important 
test of 
the $RMF$ parametrizations before applying them to the dynamics 
of heavy ion collisions. 
However, it seems difficult to determine the 
relativistic structure of the iso-vector part of the nuclear mean field
 in the static case, 
because 
the influence of the $\delta$-meson is only moderate, except if one 
goes to extremely asymmetric neutron-rich nuclei. 
However, this is not the aim of the paper, but the behavior of 
the symmetry energy 
at supra-normal densities. Since in heavy ion collisions highly compressed 
matter is formed, at least for short time scales, we study now the 
dynamical case in more detail.
The high velocity fields will also enhance the different Lorentz properties
of the various contributions, \cite{grefl}.

\section{Heavy ion collisions (dynamical case)}
The theoretical treatment of heavy ion collisions at relativistic energies 
between $0.1$ and $1-2~AGeV$ is based here on a covariant transport theory. 
As detailed described in Ref. \cite{horror}, starting from a covariant quantum 
description of the nuclear many body problem one arrives at a 
transport equation which is called in the literature  
Relativistic-Boltzmann-Uhlenbeck-Uehling ($RBUU$) equation 
\begin{eqnarray}
& & \left[ 
k^{*\mu} \partial_{\mu}^{x} + 
\left( 
  k^{*}_{\nu} F^{\mu\nu} + m^{*} \partial_{x}^{\mu} m^{*}  
\right) 
\partial_{\mu}^{k^{*}} 
\right] f(x,k^{*}) = 
\nonumber\\
& & =  \frac{1}{2} \int 
\frac{d^3 k_{2}}{E^{*}_{k_{2}}(2\pi)^3} \frac{d^3 k_{3}}{E^{*}_{k_{3}}(2\pi)^3}
             \frac{d^3 k_{4}}{E^{*}_{k_{4}}(2\pi)^3} W(kk_2|k_3 k_4)   
(2\pi)^4 \delta^4 \left(k + k_{2} -k_{3} - k_{4} \right) 
\nonumber\\  
& & \times \; \Big[ \: f(x,{\bf k}_3) f(x,{\bf k}_4) 
                      \ls 1-  f(x,{\bf k}) \rs \ls 1- f(x,{\bf k}_2) \rs - 
\nonumber \\  
& & \; \; \; \; \; \:
   f(x,{\bf k}) f(x,{\bf k}_2) 
                \ls 1-  f(x,{\bf k}_3) \rs \ls 1- f(x,{\bf k}_4) \rs \: \Big] 
\label{rbuu} 
\quad ,
\end{eqnarray}

with effective masses $m^{*}$, effective momenta $k^{*\mu}$ and the 
field tensor defined by 
$F^{\mu\nu} = \partial_{x}^{\nu}\Sigma^{\mu} - \partial_{x}^{\mu}
\Sigma^{\nu}$. $\Sigma^{\mu}$ are the vector self-energies and 
$W(kk_2|k_3 k_4)$ the transition rate for the process
$k + k_{2} \rightarrow k_{3} + k_{4}$.

The main approximations involved in Eq. (\ref{rbuu}) are a semi-classical 
gradient expansion and a quasi-particle approximation which sets the effective 
momenta on shell, i.e. $k^{*\mu}k^{*}_{\mu}=m^{*2}$. 
The numerical procedure is performed within the relativistic Landau Vlasov 
method \cite{rlv,gait96}, where the phase space distribution function 
$f(x,k^{*})$ 
is characterized by test particles of a covariant Gaussian shape in 
coordinate and momentum space. Sufficient stability is achieved 
when using $\sim 40-50$ test particles per nucleon for the simulations.

Protons and neutrons are propagating separately according to their
hadronic fields and the Coulomb interaction. 
The collision integral is treated by Monte-Carlo methods \cite{bertsch} 
including all important inelastic channels up to $1-2~AGeV$ \cite{coll}. 
Here we explicitely propagate 
$\Delta^{0,\pm,++}-$ and $N^{*}$-resonances including their decay to one- and 
two-pion ($\pi^{0,\pm}$) channels. 
At these energies we consider the lower mass resonances only,
 with free mass assignements $M_{\Delta}=1232~MeV$ and $M_{N^*}=1440~MeV$.
For elastic scattering we apply the 
empirical evaluations of isospin dependent cross sections \cite{cugnon}. 
For the inelastic channels we use the free cross section
parametrizations of ref. \cite{aichelin}. 
The produced pions are also
propagated under the influence of the Coulomb field. 
We should mention that the parameters in the collision 
term are commonly used by other groups. 
Main properties as rapidity distributions 
and transverse momentum spectra of protons as well as pion 
multiplicities are in agreement with results from other groups when 
using similar parameters for the nuclear equation of state 
\cite{lar01,trento}. 

It has been suggested since some time that the difference in neutron-proton
flows should be a sensitive observable to study the density behavior
of the isovector $EOS$. Several calculations in the non-relativistic
framework have indeed shown such sensitivity to the asymmetry-stiffness,
 but rather extreme choices were involved, see refs. \cite{bao00,sca99}.
In a previous letter \cite{grefl} we have investigated proton-neutron
differential flows in the framework of the present $RMF$ models. Also in this 
case a strong sensitivity was found, at high energies, clearly due to
the Lorentz structure of the isovector interaction. However no 
data exist at present. In this work we will concentrate on pion production 
and spectra. We are also motivated by the fact that 
recent experimental results on isospin effects in pion production 
are starting to appear, \cite{fopi,kaos}.

\begin{figure}[t]
\unitlength1cm
\begin{picture}(10.,9.0)
\put(0.0,0.0){\makebox{\epsfig{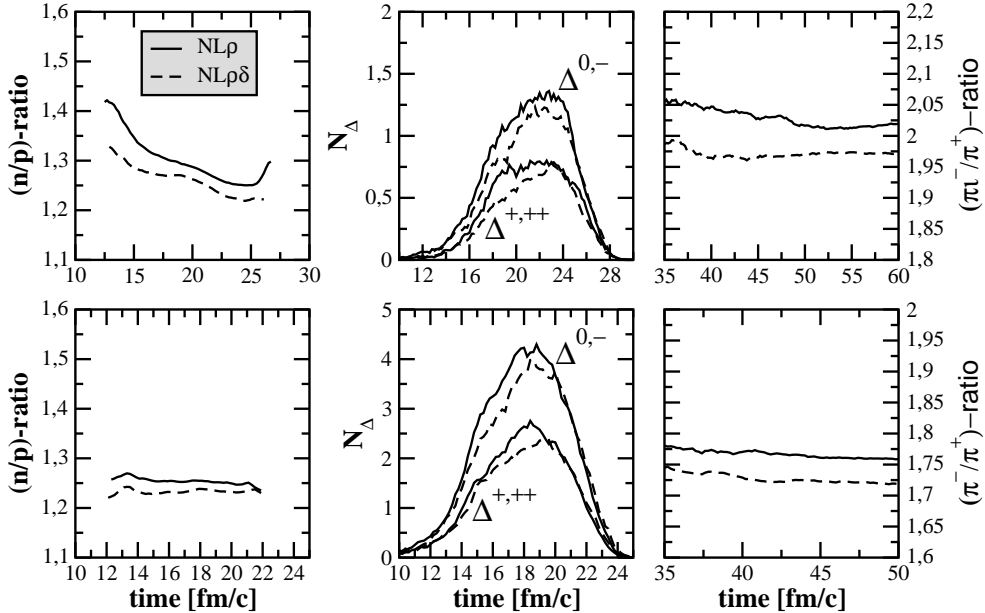}}}
\end{picture}
\caption{Time dependence of the ($\frac{n}{p}$)-ratio (left), the
multiplicities of 
$\Delta^{0,-},~\Delta^{+,++}$ resonances (middle) and 
the ($\pi^{-}/\pi^{+}$)-ratio (right) for central ($b<2~fm$) $Au+Au$ 
collisions at $0.6$ (top) and $1~AGeV$ (bottom). Results of 
$RBUU$ calculations 
with the $\rho$-meson ($NL\rho$, solid) only and both, the $\rho$ 
and $\delta$ mesons 
($NL\rho\delta$, dashed) for the isovector channel are shown. 
}
\label{fig6}
\end{figure}

\subsection{Neutron/proton densities and $\pi^-/\pi^+$ ratios}

Observable effects originating from the high density symmetry energy 
should be related to differences on ratios of neutron and proton 
densities. Thus, we consider in the following the ratio of neutrons to protons 
as a function of time and space. One also expects isospin effects on different 
isospin channels of pions, since they are produced via $nn$, $pp$ 
and $np$ collisions 
through the decay of $\Delta$ and $N^{*}$ resonances.

Fig.~\ref{fig6} shows the time evolution of the ratio of neutrons to 
protons for regions of density higher than saturation, of the 
multiplicities of $\Delta^{0,-}$- and 
$\Delta^{+,++}$-resonances and finally of the ratio of negative to positive 
pions for central (impact parameter $b<2~fm$) $Au+Au$ collisions at $0.6$ 
and $1.0~AGeV$ using 
the $NL$ models without (solid) and with (dashed) the $\delta$ field.
In the $NL\rho\delta$ calculations the neutrons are emitted much earlier
than protons from the high density phase due to a more repulsive mean field
joined to a lower $n$-effective mass. This 
effect is responsible for the reduction of the $n/p$-ratio in the
residual system and, particularly, it 
influences the particle production, see also \cite{bali,uma}.
 
The four isospin states of the $\Delta$-resonance are produced from 
different scattering channels, when the threshold energy is available, e.g. 
$nn \rightarrow p\Delta^{-},n\Delta^{0},~pp \rightarrow p\Delta^{+},n\Delta
^{++},\cdots$. Thus
$\Delta^{0,-}$ ($\Delta^{+,++}$) resonances are mainly formed in energetic 
$nn$- ($pp$-) collisions. Therefore, the sensitivity on 
the density dependence of the symmetry energy  of the $n/p$-ratio
 is indirectly 
related to that of the particle production. 

The $n/p$ effective mass
splitting in asymmetric matter can also directly affect the resonance 
production 
around the threshold, with noticeable effects on the yields. We remind 
that resonances will also have isospin dependent in-medium effective masses
that can be roughly related to the nucleon effective masses just using
the isopin coupling coefficients in the process $\Delta \leftrightarrow \pi{N}$
\cite{bali}. Thus the effective masses of the four isospin states of the
$\Delta$-resonance will be different when the $\delta$-meson is included
in the calculation. In any case we know   
from
general grounds, see Section 2, that, in $n$-rich systems, a 
isovector scalar meson field
is leading to a neutron effective mass smaller than proton, in particular 
at high baryon densities. Consequently in the process
$nn \rightarrow p\Delta^{-}$ less energy will be available for the
 $\Delta^-$ production in the $NL\rho\delta$ case.

We like to note that this mechanism is not necessarily linked to the 
density behavior
of the symmetry energy. E.g., in density dependent coupling models we
can have at high densities a soft $E_{sym}(\rho)$ just because the
$\rho$-meson coupling is decreasing but still a large $m^*_p-m^*_n$
splitting if the $\delta$-meson coupling stays constant or slightly
increasing, like in the $DBHF$ estimations (see Fig.\ref{fig1}).

In the resonance multiplicity  
(middle 
figures) one thus sees a decrease of the $\Delta^{0,-}$ isospin 
states due to the 
effect of the isovector-scalar $\delta$-meson. The pions, on the other hand,
 are mainly produced 
from resonance decay according, e.g. 
$\Delta^{-} \rightarrow n \pi^{-},~\Delta^{0} \rightarrow p\pi^{-},
~\Delta^{+} \rightarrow n\pi^{+}, \cdots$. Therefore the decrease of 
$\Delta^{0,-}$ resonances reduces 
the production of negative charged pions which consequently decreases 
the $\pi^{-}/\pi^{+}$-ratio 
(right part in Fig.~\ref{fig6}). This ratio then appears to be sensitive 
to the isospin term of equation of state at high densities. However 
it is also affected by secondary pion absorption channels which 
start to influence the results at high energies. 

One should note that 
both $NL$ models used here exhibit a similar asy-stiff behavior at 
high densities (see Fig.\ref{fig2}). We can deduce that the
effective $n/p$ mass splitting mechanism described before
is very important in inducing isospin effects on the particle production
\cite{masses}.

\begin{figure}[t]
\unitlength1cm
\begin{picture}(10.,9.0)
\put(0.0,0.0){\makebox{\epsfig{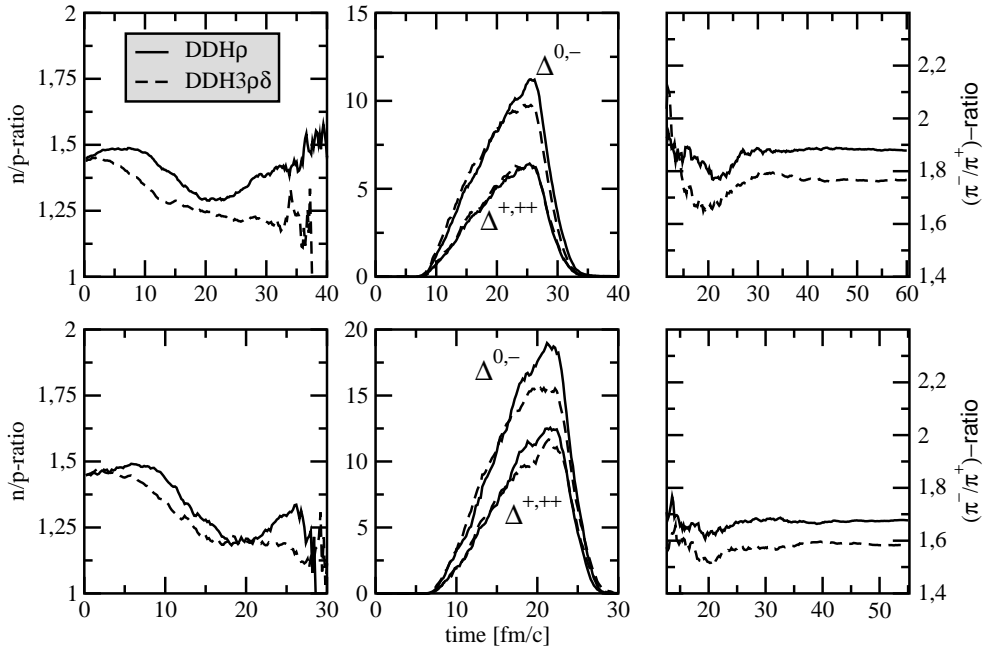}}}
\end{picture}
\caption{Same as in Fig.~\ref{fig6}, but with the $DDH\rho$ (solid) and
$DDH3\rho\delta$ (dashed) models where the 
isovector couplings are explicitely density dependent. Note the larger
$y-$scales compared to the $NL$ cases of Fig.~\ref{fig6}.
}
\label{fig7}
\end{figure}
Next Fig. \ref{fig7} shows the same quantities as in Fig.\ref{fig6} but 
now using the 
$DDH$ models where 
all coupling functions are density dependent.
In this case we have compared the $DDH\rho$, very soft symmetry energy,
with $DDH3\rho\delta$, relatively stiff one.
Due to the strong density decrease of the $\rho-$ coupling function in the
$DDH\rho$ parametrization,
now the influence of the introduction of the iso-vector $\delta$ channel 
is more pronounced comparing to the $NL$ cases, because of the 
larger differences in the 
high density symmetry energy and in the $n/p$ effective mass
splitting. In the transport 
calculations with the $\delta$ meson (left-center panels of Fig.\ref{fig7})
the $n/p$-ratio and the $\Delta^-$-yield decrease  at 
high density 
is indeed stronger than in the $NL$ models (Fig.\ref{fig6}). 
Consequently we also see a stronger influence on pion production. 

By comparing the Figs.\ref{fig6} and \ref{fig7} we see clear differences 
in the 
total multiplicity of resonances and therefore in the 
absolute pion yields.
 One should note that the $DDH$ parametrizations differ from the 
$NL$ models also in the iso-scalar part of the mean field. In particular, 
the $EOS$ for 
symmetric nuclear matter, which is similar for both models around saturation,
has a smaller incompressibility at $\rho_0$ in the $DDH$ cases
(see the discussion in Sect.2) and it is also softer at high densities 
 \cite{lenske,typel}. 
Thus, in the dynamic calculations with the $DDH$ parametrizations one 
naturally obtains more 
compression in the central region of the reaction which, in turn, 
is responsible 
for the production of higher multiplicities of resonances.
The factor $5-7$ difference in the $\Delta$ formation probability
during the compression phase, see the middle panels in Figs.\ref{fig6}-
\ref{fig7}, seem to indicate that pion absolute yields can be a good probe
of the nuclear $EOS$ at high densities, apart from corrections due to
in-medium modifications of inelastic cross sections, resonance decays 
 and $\pi$-reabsorption effects, see ref.\cite{lar01}. 

The comparison 
between 
the $DDH$ and $NL$ models with respect to isospin effects depends on
several factors. However, the reason of the presentation of these different 
models 
was to focus on the 
effects originating from the introduction of the scalar 
iso-vector $\delta$ field 
in the symmetry energy.  We have shown that there are common features
in all models, namely a larger repulsion and a smaller effective
mass for neutrons in high baryon and isospin density regions, for
$n$-rich systems.
A decrease of the $n/p$ ratios and of the $\Delta^{-,0}$ multiplicities
will finally reduce the charged pion $\pi^-/\pi^+$ ratios. 

\begin{figure}[t]
\unitlength1cm
\begin{picture}(10.,9.0)
\put(0.0,0.0){\makebox{\epsfig{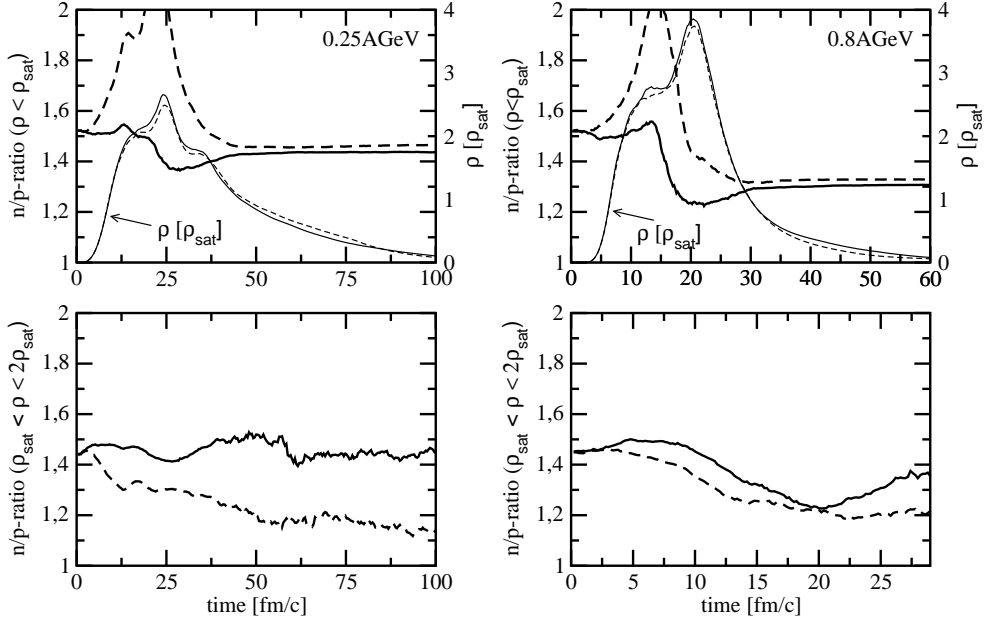}}}
\end{picture}
\caption{Time evolution of the $n/p$-ratio (left scale) for  
density regions below (upper panels) and above (lower panels)
saturation density in $Au+Au$ collisions at energies of $0.25$
(left panels) and $0.8~AGeV$ (right panels).
The upper panels also show the evolution of the central total densities
in units of $\rho_{sat}$ (thin lines, right hand scale).
The models shown are the same as in Fig.\ref{fig7}: $DDH-\rho$ (solid lines)
 and $DDH3\rho\delta$ (dashed). 
}
\label{fig8}
\end{figure}

\subsection{Isospin diffusion}
 
Before we continue to discuss the pion yields in more detail, we briefly
make some further comments on the $n/p$ ratios in different density regions. 
This is related to the isospin diffusion from high to low density regions
and with the subsequent formation of fragments. This has been suggested in 
various works \cite{bar02,kofra} as another means of investigating
the density dependence of the isovector $EOS$. 
In Fig.\ref{fig8} we show the time evolution of the $n/p$ ratio in regions 
with density below (upper panels, ``gas and light clusters'') and above
(lower panels, ``liquid'') saturation density for two incident energies.
In the upper panels we also show the total central baryon density
as an indicator of the global evolution of the collision. We use the
extreme density dependent coupling models as in Fig.\ref{fig7}, namely
the very asy-soft $DDH\rho$ (solid) and the asy-stiff $DDH3\rho\delta$
(dashed) interactions.

One sees in the upper panels that there is a strong emission of neutrons
to the gas just before and at maximum compression for the asy-stiff model,
where there is a larger differential driving force on the neutrons.
The effect is essentially non existing for the asy-soft model.
In correspondence the $n/p$ ratio of the liquid (lower panels) 
decreases below the initial 
value, more so for lower energies.
Consequently the chemical composition of the light clusters 
rapidly formed during the expansion phase will show signatures 
of the asy-stiffness of the $EOS$, as already suggested in ref.\cite{kofra}.
However, at later times the $n/p$ ratio of the dilute region is very similar
for the two interactions (slightly below the initial value), meaning that 
at these energies the bulk of
the matter finally ends up in free nucleons and light clusters.


\subsection{Energy dependence of $\pi^-/\pi^+$-ratios}

We now take up a more detailed discussion of isospin effects on pion yields.
It was see in Figs.\ref{fig6},\ref{fig7} that the $\pi^-/\pi^+$-ratios
stabilize after the compression phase and that they depend on the isovector
$EOS$ and, particularly, on the inclusion of the $\delta$-field. The latter
is generally leading to a more $asy-stiff~EOS$ and in any case to a
reduction of the neutron effective mass vs. the proton one. 
The larger neutron repulsion 
produces less neutron rich matter, which in turn produces less negative 
charged $\Delta$-resonances. The same effect is directly reached by
the reduction of the neutron effective mass. Correspondly less
negative pions are formed, leading to a reduction of the
$\pi^-/\pi^+$-ratio of the order of $5~to~10\%$.

\begin{figure}[t]
\unitlength1cm
\begin{picture}(10.,10.0)
\put(0.0,0.0){\makebox{\epsfig{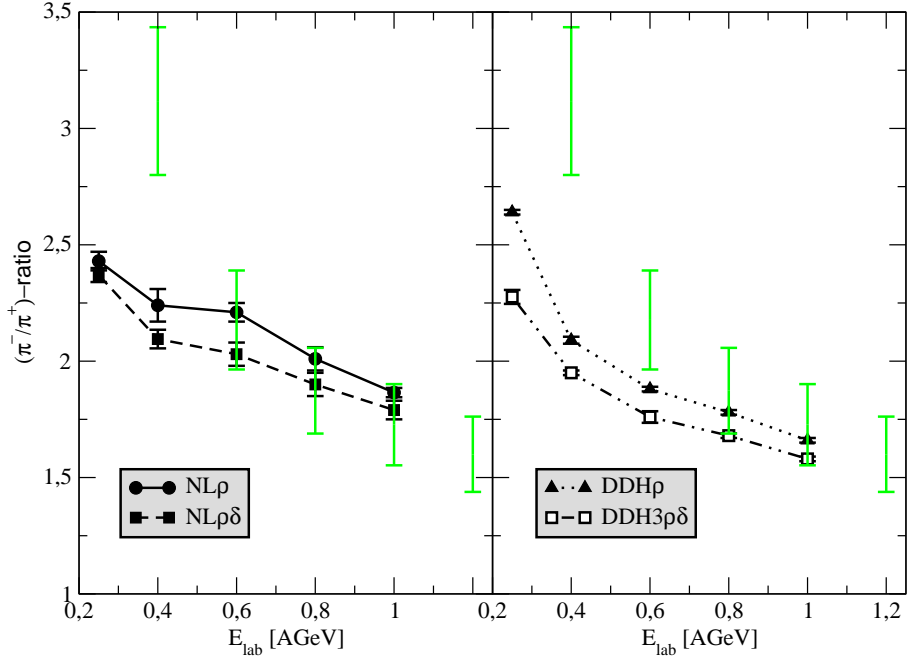}}}
\end{picture}
\caption{Energy dependence of the ($\pi^{-}/\pi^{+}$)-ratio in central
$Au+Au$ collisions. Left panel: $NL$ results, $NL\rho$ (solid) and
$NL\rho\delta$ (dashed). Right panel: $DDH$ results, $DDH\rho$ (solid)
and $DDH3\rho\delta$ (dashed). The grey bars correspond to some very
pèreliminary data from the $FOPI-GSI$ collaboration \cite{fopi}.
}
\label{fig9}
\end{figure}

All the isospin effects on pion production are generally decreasing
with incident energy, see Fig.\ref{fig9}, due to secondary collisions,
i.e. pion absorption and $\Delta$-rescattering. 
Thus all the isospin dependence
of pion production will be on average moderated.
Such effect was already found in earlier studies 
using a non-relativistic Quantum-Molecular Dynamics ($QMD$) 
approach \cite{uma}.

The beam energy behavior of the $\pi^-/\pi^+$-ratio 
evaluated at the $freeze-out$
time for $Au+Au$ central reactions are shown in Fig.\ref{fig9},
 for the $NL$ models (left panel) and for the $DDH$ models (right panel).
We like to note that in both cases the $\delta$ influence, difference between
solid and dashed curves, is decreasing with the beam energy. This is 
a nice indication of the $n/p$ effective mass splitting mechanism
in $\Delta^-$ production, expected
to be more important at the threshold.  

For the same systems new very preliminary data on pion production 
have recently been reported from the $FOPI$-collaboration 
at $GSI$ \cite{fopi}, shown as grey bars in the 
Fig.\ref{fig9}. These experimental indications are used to test
our miscroscopic evaluations.
 
We see that in our simulations we generally have a qualitative 
good description of the 
pion ratio with respect to beam energy, except perhaps at the lower energies. 
The inclusion of a
$\delta$ meson in the iso-vector part of the equation of state seems to
improve the comparison, at least at higher energies.
In any case it appears that the $\pi^-/\pi^+$-ratios might be a
probe to the role of the high density symmetry energy and the role of the
virtual $\delta$-meson. It would be nice to look at more exclusive data
on pion production in asymmetric systems. This will be the subject of the 
next section on pion flows.
\begin{figure}[t]
\unitlength1cm
\begin{picture}(10.,9.0)
\put(0.0,0.0){\makebox{\epsfig{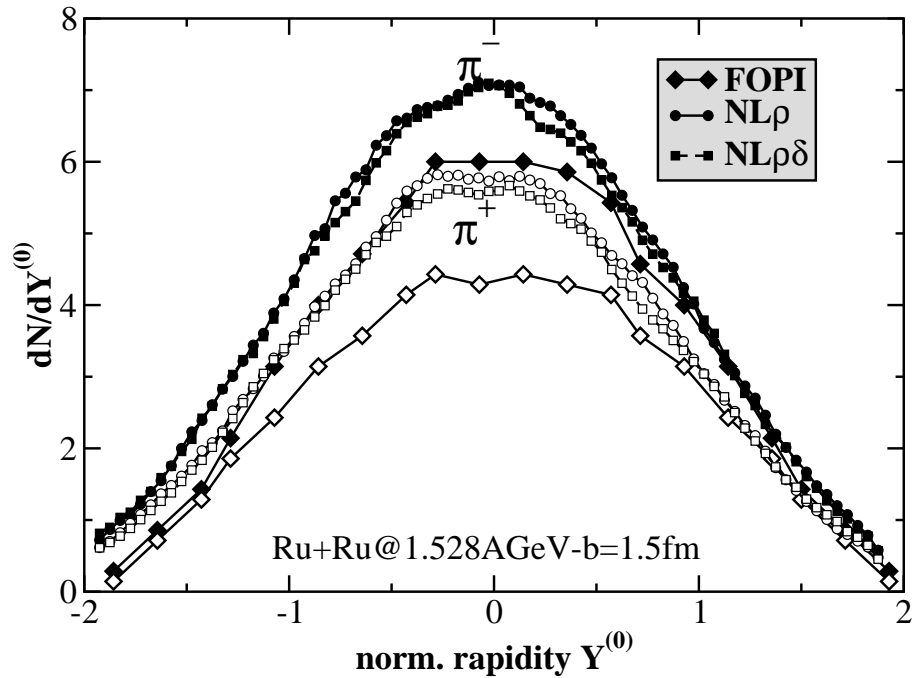}}}
\end{picture}
\caption{$\pi^{\pm}$-rapidity distributions for central $^{96}Ru+^{96}Ru$
 collisions 
with $NL\rho$ and $NL\rho\delta$ models, 
without the Coulomb field
for pion propagation. Full symbols are for $\pi^-$, open ones for $\pi^+$.
Data are taken from Ref.~\protect\cite{hong}. 
}
\label{fig10}
\end{figure}

\section{Isospin effects on pion flows}

We now discuss in more detail the velocity distributions of pions,
i.e. flow results. In spite of the relatively low asymmetry of the
system ($N/Z=1.18$) we will study central $^{96}Ru+^{96}Ru$ collisions
at about $1.5~AGeV$, since recent data are available from the
$FOPI$ collaboration \cite{hong}. We will show symmetry energy effects
and how in this respect the Coulomb interaction in pion propagation
is important. In particular we can expect the role of the Coulomb field 
to be not negligible in high density regions, for the 
interaction between the 
produced pions and the charged hadronic enviroment. 
We will also see that without taking the Coulomb effects into account, 
the differences originating 
from the $\delta$ meson are largely reduced. 
 
\begin{figure}[t]
\unitlength1cm
\begin{picture}(10.,9.5)
\put(0.0,0.0){\makebox{\epsfig{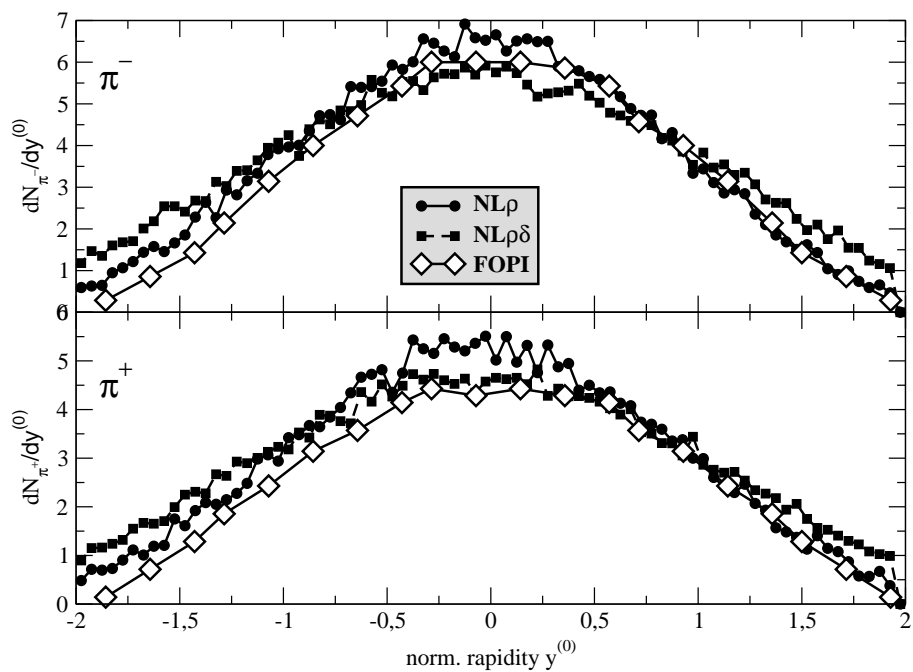}}}
\end{picture}
\caption{$\pi^{\pm}$-rapidity distributions for the same system 
as in Fig.~\protect\ref{fig10}, for the $NL\rho$ (full circles)
and $NL\rho\delta$ (full squares) models, but including the 
Coulomb field in 
the pion propagation. Upper panel: $\pi^-$. Lower panel: $\pi^+$.
Data (open diamonds) are taken from Ref. \protect\cite{hong}. 
}
\label{fig11}
\end{figure}

In Figs.\ref{fig10} and \ref{fig11} we show the rapidity distributions
for $\pi^{\pm}$ in comparison to the data for the $NL\rho$ and $NL\rho\delta$
models. The normalized rapidity is defined 
by $Y^{0} \equiv (Y/Y_{proj})_{cm}$. 

In the calculations of Fig.\ref{fig10} the Coulomb interaction was
turned off for the pion propagation. It is seen that without the Coulomb 
interaction the rapidity distribution is generally too high compared
to experiment, particularly so at mid-rapidity.  
Including the Coulomb, Fig.\ref{fig11}, the distributions are reduced,
in much better agreement with the data. The effect of including
$\delta$-fields is also enhanced.

During the high compression stage the Coulomb field accelerates the 
positively charged pions away from mid-rapidity. The opposite is happening
with the $\pi^-$'s that will be attracted towards the higher density region
where they will suffer a larger absorption.
This leads to a broadening
of the pion distributions and to a general reduction due to the
enhanced rescattering and absorption. Similar effects were also seen
in the non-relativistic calculations of ref.\cite{uma}.

The combined role of the $\delta$- and Coulomb- fields on 
the difference between positively 
and negatively charged pions is a more subtle effect. 
As we have seen previously
the structure of the symmetry term affects the isospin diffusion from 
high to low density regions. Moreover the $\delta$ contribution is directly
reducing the negative $\Delta$-resonance production through the $n/p$-mass
splitting. As a consequence the $NL\rho\delta$ model leads to a more 
positively charged hadronic central region and to larger Coulomb effects,
The freeze-out $\pi^{-}/\pi^{+}$ ratio at mid rapidity is then reduced.
However, the Coulomb field, 
acting more repulsive, is inducing larger reabsorption effects for 
high momentum $\pi^{+}$. As a consequence we can have an
enhancement of the $\pi^{-}/\pi^{+}$ ratio for high-rapidity pions. 
The influence of 
the $\delta$-channel is thus observable in terms of pion stopping. 

This can be seen in Fig.\ref{fig12} (left panel), where we show
the rapidity dependence of the $\pi^-/\pi^+$-ratio for the two $NL$ models.
For mid-rapidity the inclusion of the $\delta$-field reduces the ratio
with the mechanisms described before. Since the mid-rapidity pions are the 
most abundant (Fig.\ref{fig11}), this is consistent with the global reduction 
of the $\pi^-/\pi^+$-ratio 
(see Fig.\ref{fig9}, left panel).

\subsection{Transverse momentum dependence}

All the effects discussed before should be particularly important 
in the compressional
phase of the collision, while pions are produced all over the reaction 
evolution up to the freeze-out. A good selection procedure will be to
look at particles with high transverse momenta $P_t$. It is indeed physically
intuitive, and confirmed by simulations \cite{grefl,uma,gait01}, that 
particles emitted from the earlier high density phase have higher $P_t$'s.

\begin{figure}[t]
\unitlength1cm
\begin{picture}(10.,9.0)
\put(0.0,0.0){\makebox{\epsfig{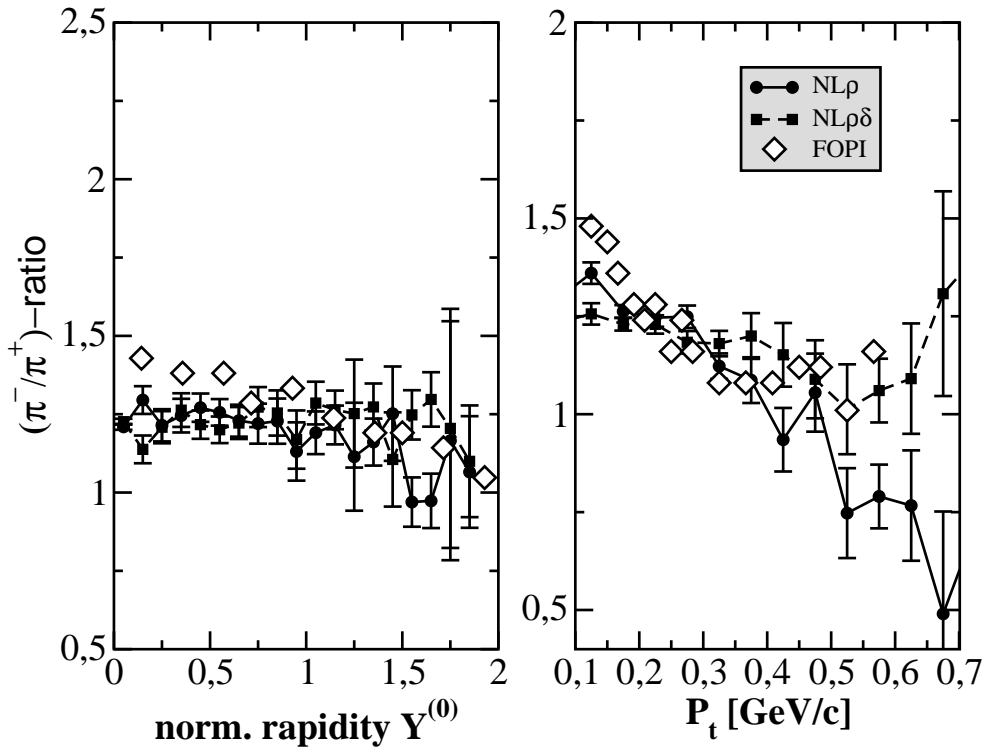}}}
\end{picture}
\caption{$\pi^{-}/\pi^{+}$-ratio for central $Ru+Ru$-collisions
as a function of rapidity (left) and transverse momentum (right) 
at mid-rapidity ($\vert Y^{(0)} \vert \leq 0.15$). 
Same models as in Fig.~\ref{fig10}. Data are taken from Ref. 
\protect\cite{hong}. 
}
\label{fig12}
\end{figure}

The $P_t$-dependence of the $\pi^-/\pi^+$-ratio in central $Ru+Ru$ collisions
for mid-rapidity pions is shown in the right panel of Fig.\ref{fig12}.
It is seen that the inclusion of the $\delta$-field strongly enhances
the ratio for high-$P_t$ pions, bringing it into more agreement with
the data. 

It is here that one sees the combined symmetry and Coulomb effects most 
clearly. With inclusion of the $\delta$-field the high density phase
is relatively more rich in protons and positively charged resonances.
The Coulomb force accelerates $\pi^+$ away from this region and 
so reduces their number at mid-rapidity. All that then
enhances the $\pi^-/\pi^+$-ratio for high-$P_t$ 
particles. 

We have repeated the analysis for the two $DDH$ interactions already
studied before, $DDH\rho$ and $DDH3\rho\delta$. The inclusion of
the $\delta$-channel leads to the same effect. We see an increase
of the $\pi^-/\pi^+$-ratio for high-$P_t$ pions at mid-rapidity,
as observed in experiments.

We have shown that $\pi^{-}/\pi^{+}$-ratios
for high transverse momentum pions are a sensitive probe
to the structure of the symmetry term, in particular to the contribution
of a $\delta$-field. The trend of the data with rapidity and transverse
momentum is affected by the dynamical effects of the inclusion of a
isovector scalar field in the hadronic model. This seems in better
agreement with the data, in particular for high-$P_t$ emitted pions.

We finally remark that this already clean indication could be
further supported by studying instead of $^{96}Ru+^{96}Ru$ ($N/Z=1.18$)
more asymmetric cases, like $^{132}Sn+^{124}Sn$ ($N/Z=1.56$)
or $^{197}Au+^{197}Au$ ($N/Z=1.49$) colliding systems, and
looking at similar more exclusive data.

\section{Summary and outlook}
In this paper we investigate in detail the density and momentum 
dependence of 
the symmetry energy in nuclear matter properties, finite nuclei and, 
in particular, 
in dynamical situations of intermediate energy heavy ion collisions. We 
discuss isospin effects in terms of the Lorentz structure of the 
isovector mean field. In addition to the usual vector-isovector
$\rho$-meson we introduce a scalar, isovector $\delta$-like field, 
naturally keeping the same bulk asymmetry parameter 
at saturation. This will, however, change the high density and high momentum
behavior of the symmetry term.
We study the effects 
arising from the $\delta$-meson using different models of the
Relativistic Mean Field ($RMF$) theory, then
stressing the role of Lorentz covariant properties of the 
isovector mean field in a general way. 

The comparisons between Non-Linear Walecka type ($NL$, constant couplings) and 
the microscopically motivated density dependent hadron field 
($DDH$, density dependent couplings) models show that the effects of 
the $\delta$ meson are of general 
nature and that they originate from the different covariant properties 
of the vector $\rho-$ and scalar $\delta-$fields 
at high baryon densities.

We have applied the models of asymmetric nuclear matter to finite nuclei 
and heavy ion 
collisions. 

The studies of nuclear structure show that the effects of including 
the isovector $\delta$-channel are very small, except if one goes much 
beyond the 
valley of stability near the neutron drip line. 
Observable effects are found for very neutron-rich nuclei as an increased 
neutron diffusion at the nucleus surface and a larger
binding energy 
per nucleon. Both results are due to the fact that in $n-$rich systems
at low densities neutrons feel a larger  
attractive scalar 
self energy when including the $\delta$ meson. 
However, these effects are only moderate, in agreement with similar
analyses of ref.\cite{typel}, making it 
difficult to draw definite conclusions on the effective interactions. 
The main reason relies on the observation that in finite nuclei one 
explores the $EOS$ around and below saturation density. 

Thus, in order to investigate the Lorentz structure of the high 
density symmetry 
energy we have applied the models to dynamical situations in heavy 
ion collisions at intermediate energies, where relatively high
baryon density regions can be reached during the reaction dynamics. 

The isovector scalar and vector $\delta$- and $\rho$-fields cancel
each other in a similar way as for the isoscalar $\sigma$- and 
$\omega$-fields. Thus, with the constraint to have the same
symmetry energy at saturation,  
this cancellation is less effective at higher densities, with
the repulsive $\rho$-meson gaining more importance \cite{liu02,gre03}.
With this pure relativistic mechanism the inclusion of a $\delta$-field
will make the symmetry energy more repulsive at high baryon densities.
Another very important transport effect of a scalar field in the isovector 
channel derives from the influence on the effective masses. In asymmetric
matter we have now a
splitting of the $n/p$ effective masses, also increasing with baryon density.
This will affect nucleon velocity distributions and particle
productions, particularly around the threshold.
Finally, at high relativistic energies, due to the large velocity fields, 
one can expect
large effects directly from the different covariant structure of the 
$\rho$ and $\delta$ mesons \cite{grefl}.

We have studied heavy ion collisions in a relativistic 
transport model of 
Boltzmann type and analyzed the results in terms of neutron and 
proton densities in
space and time, particle production and pion flow observables.
Within the same frame we have studied nucleon flows in a previous letter
\cite{grefl}.

For $n$-rich colliding systems we find a 
strong decrease 
of the $n/p$-ratio in the bulk matter when including the 
$\delta$ meson in the 
isovector mean field. We interpret this effect by the joint action
of a stiffer symmetry term and   
of a decrease of the 
neutron effective mass, more pronounced at high baryon densities, 
which both lead to more repulsion for neutrons than for protons. 
The influence of the 
$\delta$-meson in the dynamical situations has been mainly studied 
within the $NL$-models, 
where the differences in high density symmetry energy are actually not so 
strong. 
However, the $DDH$ models of density dependent 
coupling functions 
show very similar effects, even more pronounced due to the 
softer character of the 
symmetry energy. Moreover the dynamical implications of nucleon effective 
mass splitting are present in all cases.
Therefore, the influence of the $\delta$-channel is a general 
feature arising from the differences in the
 Lorentz structure of 
the high density symmetry energy. 

The relativistic form of the 
high density isovector 
$EOS$ can also be directly seen in terms of particle production which can 
be experimentally measured. For this reason we have focussed our
study on the influence 
of the $\delta$-meson on pion production. We evaluate from the 
reaction simulations the ratios of 
charged pions as function of time and space, energy, rapidity and transverse 
momentum. In general for all the colliding systems and energies we observe 
a reduction of the $(\pi^{-}/\pi^{+})$-ratio in the models 
containing the $\delta$-meson. 

This is due to the fact that the $\delta$ meson makes the high density 
region more proton-rich, see before. This induces a reduction in the 
formation of $\Delta^{0,-}$ resonances and thus in the production 
of negative charged pions. 
Moreover the lower neutron effective masses represent an efficient
mechanism to directly quench the formation of negative 
$\Delta$-resonances. This is particularly evident at lower energies, 
around the threshold for resonance production.

The effect of the $\delta$-field as discussed above is enhanced in the 
momentum spectra, i.e. the rapidity and transverse momentum distributions,
due to the interplay with the Coulomb interaction which affects differently
the $\pi^+$ and $\pi^-$ distributions. Thus a clear, and likely detectable,
effect of the $\delta$-field is predicted in the flow observables of
the $(\pi^{-}/\pi^{+})$-ratios, as well as in the incident energy dependence.

The comparison with 
the experiments seems to indicate that the introduction of 
the $\delta$ meson in the 
isovector mean field is important, i.e. as a way to promote, in
$n-$rich systems,
a larger vector repulsive field for high energy neutrons, joint
to a smaller effective mass. We remark that both effects are mainly 
due to the relativistic dynamics and not necessarily directly
related to the high density behavior of the symmetry energy. 

However, in order to strengthen these conclusions
one needs 
more precise experimental data, with more asymmetric, even unstable,
ion beams, not only on pion production but 
simultanously on 
collective flow of neutrons {\it and} protons to study the so called 
{\it isospin collective flow} \cite{grefl}. 

In the same heavy ion collisions, with high isospin density,
the threshold production of other particles and resonances would also 
provide very
relevant information.

At this level of investigation we conclude that relativistic 
studies of asymmetric 
static and dynamic nuclear matter appear very important in order
to directly probe 
the Lorentz structure of the  
isovector part of the nuclear equation of state.




\begin{thebibliography}{99}
\bibitem{npmass} We note that when the effective $\delta$-field
is included we have a splitting of the neutron-proton effective 
masses in asymmetric matter. From Eq.\ref{effmass} we see that 
in n-rich systems the neutron effective mass is smaller with a 
splitting that is increasing with baryon density (for a constant
$\delta$ coupling). We can expect 
interesting transport and threshold effects.
\bibitem{coll} For the collision term we use here a modified version of
the T\"ubingen $RQMD$ simulation code, see \cite{fuc01} and refs. therein.
\bibitem{masses}
In our framework this is a characteristic relativistic effect that should
be somewhat included in the corresponding non-relativistic models. 
The importance of this contribution was not realized so far in 
previous non-relativistic studies \cite{bali}, where no isospin effects
were included in the momentum dependent part of the effective forces. 
In this case the differences
in the used symmetry 
energy parametrizations at supra-normal densities were very 
pronounced and therefore they
resulted in very large 
effects with respect to earlier neutron emission and particle production.
\end{thebibliography}
\end{document}